\documentclass{article}
\usepackage{cite}
\usepackage{graphicx}
\usepackage{dcolumn}

\input{tcilatex}
\begin{document}

\date{}
\title{Comment on: ``Rashba coupling induced by Lorentz symmetry breaking effects''
. Ann. Phys. (Berlin) \textbf{526}, 187 (2013)}
\author{Francisco M. Fern\'{a}ndez\thanks{%
fernande@quimica.unlp.edu.ar} \\
INIFTA, DQT, Sucursal 4, C.C 16, \\
1900 La Plata, Argentina}
\maketitle

\begin{abstract}
We analyze the results of a paper on ``the arising of a Rashba-like
coupling, a Zeeman-like term and a Darwin-like term induced by Lorentz
symmetry breaking effects in the non-relativistic quantum dynamics of a
spin-1/2 neutral particle interacting with external fields''. We show that
the authors did not obtain the spectrum of the eigenvalue equation but only
one eigenvalue for a specific relationship between model parameters. In
particular, the existence of allowed cyclotron frequencies conjectured by
the authors is a mere artifact of the truncation condition used to obtain
exact solutions to the radial eigenvalue equation.
\end{abstract}

In a paper published in this journal Bakke and Belich\cite{BB13b} study
``the arising of a Rashba-like coupling, a Zeeman-like term and a
Darwin-like term induced by Lorentz symmetry breaking effects in the
non-relativistic quantum dynamics of a spin-1/2 neutral particle interacting
with external fields''. They derive an eigenvalue equation for the radial
coordinate and solve it exactly by means of the Frobenius method. This
approach leads to a three-term recurrence relation that enables the authors
to truncate the series and obtain eigenfunctions with polynomial factors.
They claim to have obtained the bound-state eigenvalues ad eigenfunctions of
the model. Since the truncation condition requires that a model parameter
depends on the quantum numbers they conclude that not all the cyclotron
frequencies are allowed. In this Comment we analyze the effect of the
truncation condition used by the authors on the physical conclusions that
they derive in their paper.

It is not our purpose to discuss the validity of the model but the way in
which the authors solve the eigenvalue equation. For this reason we do not
show the main equations displayed in their paper and restrict ourselves to
what we consider relevant. We focus in the eigenvalue equation
\begin{eqnarray}
&&R_{s}^{\prime \prime }+\frac{1}{\xi }R_{s}^{\prime }-\frac{\delta _{s}^{2}%
}{\xi ^{2}}R_{s}-\frac{\alpha }{\left( 2ma_{2}\right) ^{3/4}}\xi R_{s}-\xi
^{2}R_{s}-\frac{\tau _{s}}{\left( 2ma_{2}\right) ^{1/4}}\frac{R_{s}}{\xi }%
+WR_{s}=0,  \nonumber \\
&&\delta _{s}^{2}=\gamma _{s}^{2}+2ma_{1},\;\gamma _{s}=l+\frac{1}{2}%
(1-s),\;\tau _{s}=s\frac{gb\lambda }{4m}\gamma _{s}+\frac{gb\lambda }{8m}%
,\;\alpha =gb\lambda m,  \nonumber \\
&&\zeta =2m\left( \mathcal{E}-V_{0}\right) ,\;W=\frac{\zeta }{\left(
2ma_{2}\right) ^{1/2}},\;a_{2}=m\omega ^{2}  \label{eq:eig_eq_BB}
\end{eqnarray}
where $l=0,\pm 1,\pm 2,\ldots $, $s=\pm 1$, $m$ is a mass, $\mathcal{E}$ the
energy, $a_{1},a_{2},V_{0}$ parameters of the model potential $V(\rho
)=a_{1}\rho ^{-2}+a_{2}\rho ^{2}+V_{0}$ and $a$, $b$, and $\lambda $ are
constants that appear in the interactions included in the model. The authors
choose units such that $\hbar =c=1$ although there are rigorous ways of
deriving dimensionless equations, as well as the choice of natural units\cite
{F20}.

The authors' eigenvalue equation (\ref{eq:eig_eq_BB}) is a particular case
of
\begin{eqnarray}
\hat{L}R &=&WR,  \nonumber \\
\hat{L} &\equiv &-\frac{d^{2}}{d\xi ^{2}}-\frac{1}{\xi }\frac{d}{d\xi }+%
\frac{\gamma ^{2}}{\xi ^{2}}-\frac{a}{\xi }+b\xi +\xi ^{2},
\label{eq:eigen_eq_R}
\end{eqnarray}
where $\gamma $, $a$ and $b$ are arbitrary real numbers that have nothing to
do with the parameters in equation (\ref{eq:eig_eq_BB}). Since the behaviour
at origin is determined by the term $\gamma \xi ^{-2}$ and the behaviour at
infinity by the harmonic term $\xi ^{2}$ we conclude that there are bound
states for all $-\infty <a,b<\infty $.

By means of the ansatz
\begin{equation}
R(\xi )=\xi ^{|\gamma |}e^{-\frac{b\xi }{2}-\frac{\xi ^{2}}{2}}P(\xi
),\,P(\xi )=\sum_{j=0}^{\infty }c_{j}\xi ^{j},  \label{eq:R_series}
\end{equation}
we derive a three-term recurrence relation for the coefficients $c_{j}$:
\begin{eqnarray}
c_{j+2} &=&\frac{b\left( 2|\gamma |+2j+3\right) -2a}{2\left( j+2\right)
\left( 2|\gamma |+j+2\right) }c_{j+1}+\frac{4\left( 2|\gamma |+2j-W+2\right)
-b^{2}}{4\left( j+2\right) \left( 2|\gamma |+j+2\right) }c_{j},  \nonumber \\
j &=&-1,0,1,\ldots ,\;c_{-1}=0,\;c_{0}=1.  \label{eq:rec_rel_gen}
\end{eqnarray}

In order to obtain polynomial solutions the authors force the termination
conditions
\begin{equation}
W=W_{\gamma }^{(n)}=\frac{8\left( |\gamma |+n+1\right) -b^{2}}{4}%
,\,c_{n+1}=0,\,n=1,2,\ldots .  \label{eq:trunc_cond}
\end{equation}
Clearly, under such conditions $c_{j}=0$ for all $j>n$ and $P(\xi )$ reduces
to a polynomial of degree $n$. In this way, they obtain analytical
expressions for the eigenvalues and the radial eigenfunctions $R_{\gamma
}^{(n)}(\xi ).$ For the sake of clarity and generality we will use $\gamma $
instead of $l$ as an effective quantum number.

For example, when $n=1$ we have
\begin{eqnarray}
W_{\gamma }^{(1)} &=&\frac{8\left( |\gamma |+2\right) -b^{2}}{4},\,a_{\gamma
}^{(1,1)}=\frac{2b\left( |\gamma |+1\right) -\sqrt{b^{2}+8\left( 2|\gamma
|+1\right) }}{2},  \nonumber \\
a_{\gamma }^{(1,2)} &=&\frac{2b\left( |\gamma |+1\right) +\sqrt{%
b^{2}+8\left( 2|\gamma |+1\right) }}{2},  \label{eq:W,a,n=1}
\end{eqnarray}
or, alternatively,
\begin{eqnarray}
b_{\gamma }^{(1,1)} &=&\frac{2\left[ 2a\left( |\gamma |+1\right) -\sqrt{%
a^{2}+2\left( 2|\gamma |+3\right) \left( 2|\gamma |+1\right) ^{2}}\right] }{%
\left( 2|\gamma |+1\right) \left( 2|\gamma |+3\right) },  \nonumber \\
b_{\gamma }^{(1,2)} &=&\frac{2\left[ 2a\left( |\gamma |+1\right) +\sqrt{%
a^{2}+2\left( 2|\gamma |+3\right) \left( 2|\gamma |+1\right) ^{2}}\right] }{%
\left( 2|\gamma |+1\right) \left( 2|\gamma |+3\right) }.  \label{eq:b,n=1}
\end{eqnarray}
When $n=2$ we obtain a cubic equation for either $a$ or $b$, for example,

\begin{eqnarray}
W_{\gamma }^{(2)} &=&\frac{8\left( |\gamma |+3\right) -b^{2}}{4},  \nonumber
\\
&&4a^{3}-6a^{2}b\left( 2|\gamma |+3\right) +a\left( b^{2}\left( 12\gamma
^{2}+36|\gamma |+23\right) -16\left( 4|\gamma |+3\right) \right)   \nonumber
\\
&&-\frac{b\left( 2|\gamma |+1\right) \left( b^{2}\left( 2|\gamma |+3\right)
\left( 2|\gamma |+5\right) -16\left( 4|\gamma |+7\right) \right) }{2}=0,
\label{eq:a,b,n=2}
\end{eqnarray}
from which we obtain either $a_{\gamma }^{(2)}(b)$ or $b_{\gamma }^{(2)}(a)$%
; for example, $a_{\gamma }^{(2,1)}(b)$, $a_{\gamma }^{(2,2)}(b)$, $%
a_{\gamma }^{(2,3)}(b)$. In the general case we will have $n+1$
curves of the form $a_{\gamma }^{(n,i)}(b)$, $i=1,2,\ldots ,n+1$,
labelled in such a way that $a_{\gamma }^{(n,i)}(b)<a_{\gamma
}^{(n,i+1)}(b)$ and it can be proved that all the roots are
real\cite{CDW00,AF20}. Notice that Bakke and Belich completely
overlooked such multiplicity of roots.

It is obvious to anybody familiar with conditionally solvable (or
quasi-solvable) quantum-mechanical models (see \cite{CDW00,AF20,F20b,F20c}
and, in particular, the remarkable review \cite{T16} and references therein
for more details) that the approach just described does not produce all the
eigenvalues of the operator $\hat{L}$ for a given set of values of $\gamma $%
, $a$ and $b$ but only those states with a polynomial factor $P(\xi )$. Each
of the particular eigenvalues $W_{\gamma }^{(n)}$, $n=1,2,\ldots $
corresponds to a set of particular curves $a_{\gamma }^{(n,i)}(b)$. On the
other hand, if we solve the eigenvalue equation (\ref{eq:eigen_eq_R}) in a
proper way we obtain an infinite set of eigenvalues $W_{\nu ,\gamma }(a,b)$,
$\nu =0,1,2,\ldots $ for each set of real values of $a$, $b$ and $\gamma $.
The condition that determines these allowed values of $W$ is that the
corresponding radial eigenfunctions $R(\xi )$ are square integrable
\begin{equation}
\int_{0}^{\infty }\left| R(\xi )\right| ^{2}\xi \,d\xi <\infty .
\label{eq:bound-state_def_xi}
\end{equation}
Notice that $\nu $ is the actual radial quantum number (that labels the
eigenvalues in increasing order of magnitude), whereas $n$ is just a
positive integer that labels some particular solutions with a polynomial
factor $P(\xi )$. In other words: $n$ is a fictitious quantum number given
by the truncation condition (\ref{eq:trunc_cond}).

It should be obvious to everybody that the eigenvalue equation (\ref
{eq:eigen_eq_R}) supports bound states for all values of $a$ and $b$ and
that the truncation condition (\ref{eq:trunc_cond}) only yields some
particular solutions. Besides, according to the Hellmann-Feynman theorem\cite
{F39} the true eigenvalues $W_{\nu ,\gamma }(a,b)$ of equation (\ref
{eq:eigen_eq_R}) are decreasing functions of $a$ and increasing functions of
$b$%
\begin{equation}
\frac{\partial W}{\partial a}=-\left\langle \frac{1}{\xi }\right\rangle ,\,%
\frac{\partial W}{\partial b}=\left\langle \xi \right\rangle .
\label{eq:HFT}
\end{equation}
Therefore, for a given value of $b$ and sufficiently large values of $a$ we
expect negative values of $W$ that the truncation condition fails to
predict. It is not difficult to prove, from straightforward scaling\cite{F20}%
, that
\begin{equation}
\lim_{a\rightarrow \infty }\frac{W_{\nu ,\gamma }}{a^{2}}=-\frac{1}{\left(
2\nu +2|\gamma |+1\right) ^{2}}.  \label{eq:W_asympt}
\end{equation}
What is more, we can conjecture that the pairs $\left[ a_{\gamma
}^{(n,i)}(b),W_{\gamma }^{(n)}\right] $, $i=1,2,\ldots ,n+1$ are points on
the curves $W_{\nu ,\gamma }(a)$, $\nu =0,1,\ldots ,n$, respectively, for a
given value of $b$.

The eigenvalue equation (\ref{eq:eigen_eq_R}) cannot be solved exactly in
the general case. In order to obtain sufficiently accurate eigenvalues of
the operator $\hat{L}$ we resort to the reliable Rayleigh-Ritz variational
method that is well known to yield increasingly accurate upper bounds to all
the eigenvalues\cite{P68} (and references therein). For simplicity we choose
the basis set of non-orthogonal Gaussian functions $\left\{ u_{j}(\xi )=\xi
^{|\gamma |+j}e^{-\frac{\xi ^{2}}{2}},\;j=0,1,\ldots \right\} $ and test the
accuracy of these results by means of the powerful Riccati-Pad\'{e} method%
\cite{FMT89a}.

As a first example, we choose $n=2$, $\gamma =0$ and $b=1$ so that $%
W_{0}^{(2)}=5.75$ for the three models $\left[
a_{0}^{(2,1)}=-1.940551663,b=1\right] $, $\left[
a_{0}^{(2,2)}=1.190016441,b=1\right] $ and $\left[
a_{0}^{(2,3)}=5.250535221,b=1\right] $. The first four eigenvalues for each
of these models are
\begin{eqnarray*}
a_{0}^{(2,1)} &\rightarrow &\left\{
\begin{array}{c}
W_{0,0}=5.750000000 \\
W_{1,0}=9.894040660 \\
W_{2,0}=14.06831985 \\
W_{3,0}=18.24977457
\end{array}
\right. , \\
a_{0}^{(2,2)} &\rightarrow &\left\{
\begin{array}{c}
W_{0,0}=-0.1664353619 \\
W_{1,0}=5.750000000 \\
W_{2,0}=10.52307155 \\
W_{3,0}=15.06421047
\end{array}
\right. , \\
a_{0}^{(2,3)} &\rightarrow &\left\{
\begin{array}{c}
W_{0,0}=-27.32460313 \\
W_{1,0}=-0.5108147276 \\
W_{2,0}=5.750000000 \\
W_{3,0}=10.90599171
\end{array}
\right. .
\end{eqnarray*}
We appreciate that the eigenvalue $W_{0}^{(2)}=5.75$ coming from the
truncation condition (\ref{eq:trunc_cond}) is the lowest eigenvalue of the
first model, the second lowest eigenvalue of the second model and the third
lowest eigenvalue for the third model. The truncation condition misses all
the other eigenvalues for each of those models and for this reason it cannot
provide the spectrum of the physical model for any set of values of $\gamma $%
, $a$ and $b$ as suggested by Bakke and Belich.

In the results shown above we have chosen model parameters on the curves $%
a_{0}^{(2,i)}(b)$. In what follows we consider the case $a=2$, $b=1$ that
does not belong to any of those curves. For this set of model parameters the
first five eigenvalues are $W_{0,0}=-3.230518994$, $W_{1,0}=4.510929109$, $%
W_{2,0}=9.532275968$, $W_{3,0}=14.19728140$ and $W_{4,0}=18.70978427$. As
said above: there are square-integrable solutions (actual bound states) for
any set of real values of $a$, $b$ and $\gamma $. The obvious conclusion is
that the dependence of the frequency $\omega $ on the quantum numbers $n$, $%
l $, $s$ ($\omega _{n,l,s}$) and the consequent allowed cyclotron
frequencies conjectured by Bakke and Belich\cite{BB13b} are just artifacts
of the truncation condition (\ref{eq:trunc_cond}). Such claims are
nonsensical from a physical point of view. To be clearer, since there are
bound states for all $a$ and $b$ then there are bound states for all $\omega
$.

\begin{figure}[tbp]
\begin{center}
\includegraphics[width=9cm]{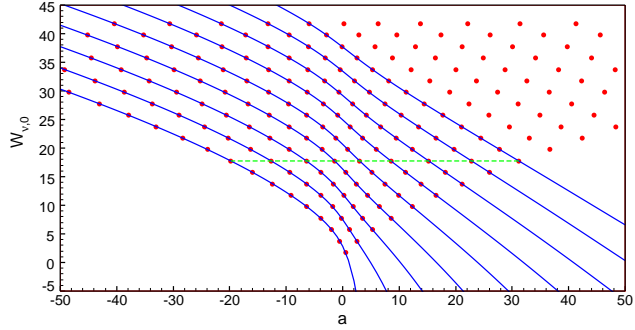}
\end{center}
\caption{Eigenvalues $W_0^{(n)}(a,1)$ from the truncation
condition (red points) and $W_{\nu,0}(a)$ obtained by means of the
variational method (blue lines)} \label{Fig:Wb1g0}
\end{figure}

Figure~\ref{Fig:Wb1g0} shows some eigenvalues $W_{0}^{(n)}(b=1)$ given by
the truncation condition (red points) and the lowest variational eigenvalues
$W_{\nu ,0}(a,1)$ (blue lines). We clearly appreciate that the truncation
condition (\ref{eq:trunc_cond}) yields only some particular points of the
curves $W_{\nu ,0}(a,1)$. Therefore, any conclusion drawn from $W_{\gamma
}^{(n)}$ is meaningless unless one is able to organize these eigenvalues
properly\cite{AF20,F20b,F20c}. Bakke and Belich\cite{BB13b} completely
overlooked this fact. The reason is that these authors appear to believe
that the only acceptable solutions to the eigenvalue equation are those with
polynomial factors $P(\xi )$. The fact is that this kind of solutions
already satisfy equation (\ref{eq:bound-state_def_xi}) but they are not the
only ones. Notice that the variational method also yields the polynomial
solutions as shown by the fact that the blue lines connect the red points in
Figure~\ref{Fig:Wb1g0}. In order to make the meaning of the eigenvalues $%
W_{\gamma }^{(n)}$ and the associated multiplicity of roots $i=1,2,\ldots
,n+1$ clearer, Figure~\ref{Fig:Wb1g0} shows an horizontal line (green,
dashed) at $W=W_{0}^{(8)}$ that intersects the curves $W_{\nu ,0}(a,1)$
exactly at the red points. The most important conclusion of present analysis
is that the occurrence of allowed oscillator frequencies are fabricated by
Bakke and Belich by picking out some isolated eigenvalues $W_{\gamma }^{(n)}$
for some particular curve $a_{\gamma }^{(n,i)}(b)$. Since there are
eigenvalues $W_{\nu ,\gamma }(a,b)$ for all real values of $a$ and $b$ then
there are bound states for every positive value of $\omega $ in their
equations (\ref{eq:eig_eq_BB}).

Summarizing: The authors make two basic, conceptual errors. The first one is
to believe that the only possible bound states are those with polynomial
factors $P(\xi )$. We have shown above that there are square-integrable
solutions for model parameters $a$ and $b$ outside the curves $a_{\gamma
}^{(n,i)}(b)$ associated to these polynomials. The second error is the
assumption that the spectrum of the problem is given by the truncation
condition (\ref{eq:trunc_cond}). It is clear that this equation only
provides one energy eigenvalue for a particular set of model parameters
given by the curves just mentioned. From these mistakes the authors
conjecture the existence of allowed cyclotron frequencies. Here we have
shown that such allowed cyclotron frequencies are fabricated by Bakke and
Belich by means of the truncation method. Therefore, such conclusion is
nonsensical from both mathematical and physical points of view. It is clear
that there are bound states for all values of $\omega $ because the
eigenvalues $W_{\nu ,\gamma }(a,b)$ are continuous functions of both $a$ and
$b$.

\begin{thebibliography}{99}
\bibitem{BB13b}  K. Bakke and H. Belich, Ann. Phys. (Berlin) \textbf{526},
187 (2013).

\bibitem{F20}  F. M. Fern\'{a}ndez, Dimensionless equations in
non-relativistic quantum mechanics, arXiv:2005.05377 [quant-ph].

\bibitem{F39}  R. P. Feynman, Phys. Rev. \textbf{56}, 340 (1939).

\bibitem{CDW00}  M. S. Child, S-H. Dong, and X-G. Wang, J. Phys. A \textbf{33%
}, 5653 (2000).

\bibitem{AF20}  P. Amore and F. M. Fern\'{a}ndez, Phys. Scr. \textbf{95},
105201 (2020). arXiv:2007.03448 [quant-ph]

\bibitem{F20b}  F. M. Fern\'{a}ndez, The rotating harmonic oscillator
revisited, arXiv:2007.11695 [quant-ph].

\bibitem{F20c}  F. M. Fern\'{a}ndez, The truncated Coulomb potential
revisited, arXiv:2008.01773 [quant-ph].

\bibitem{T16}  A. V. Turbiner, Phys. Rep. \textbf{642}, 1 (2016).
arXiv:1603.02992v2

\bibitem{P68}  F. L. Pilar, Elementary Quantum Chemistry (McGraw-Hill, New
York, 1968).

\bibitem{FMT89a}  F. M. Fern\'{a}ndez, Q. Ma, and R. H. Tipping, Phys. Rev.
A \textbf{39}, 1605 (1989).
\end{thebibliography}
\end{document}